\documentclass[journal=jacsat,manuscript=article]{achemso}
\usepackage[version=3]{mhchem} 
\usepackage{xr}
\usepackage[utf8]{inputenc}
\usepackage{graphicx}
\usepackage{hyperref}
\usepackage{caption}
\usepackage[normalem]{ulem}
\usepackage{verbatim}

\usepackage[dvipsnames]{xcolor}

\title{Exploring kinase DFG loop conformational stability with AlphaFold2-RAVE}
\author{Bodhi P. Vani}

\affiliation{Institute for Physical Science and Technology, University of Maryland, College Park, Maryland 20742, USA}
\author{Akashnathan Aranganathan}
\affiliation{Biophysics Program and Institute for Physical Science and Technology, University of Maryland, College Park 20742, USA}
\author{Pratyush Tiwary}
 \affiliation{Department of Chemistry and Biochemistry and Institute for Physical Science and Technology, University of Maryland, College Park 20742, USA}
 \altaffiliation{Corresponding author}
 \email{ptiwary@umd.edu}
\date{\today}

\abbreviations{MD,AF2,PDB}
\keywords{Protein dynamics, Free Energy, Reaction Coordinates, AlphaFold2, Molecular Dynamics}

\begin{document}

\begin{tocentry}

\includegraphics[width=\textwidth]{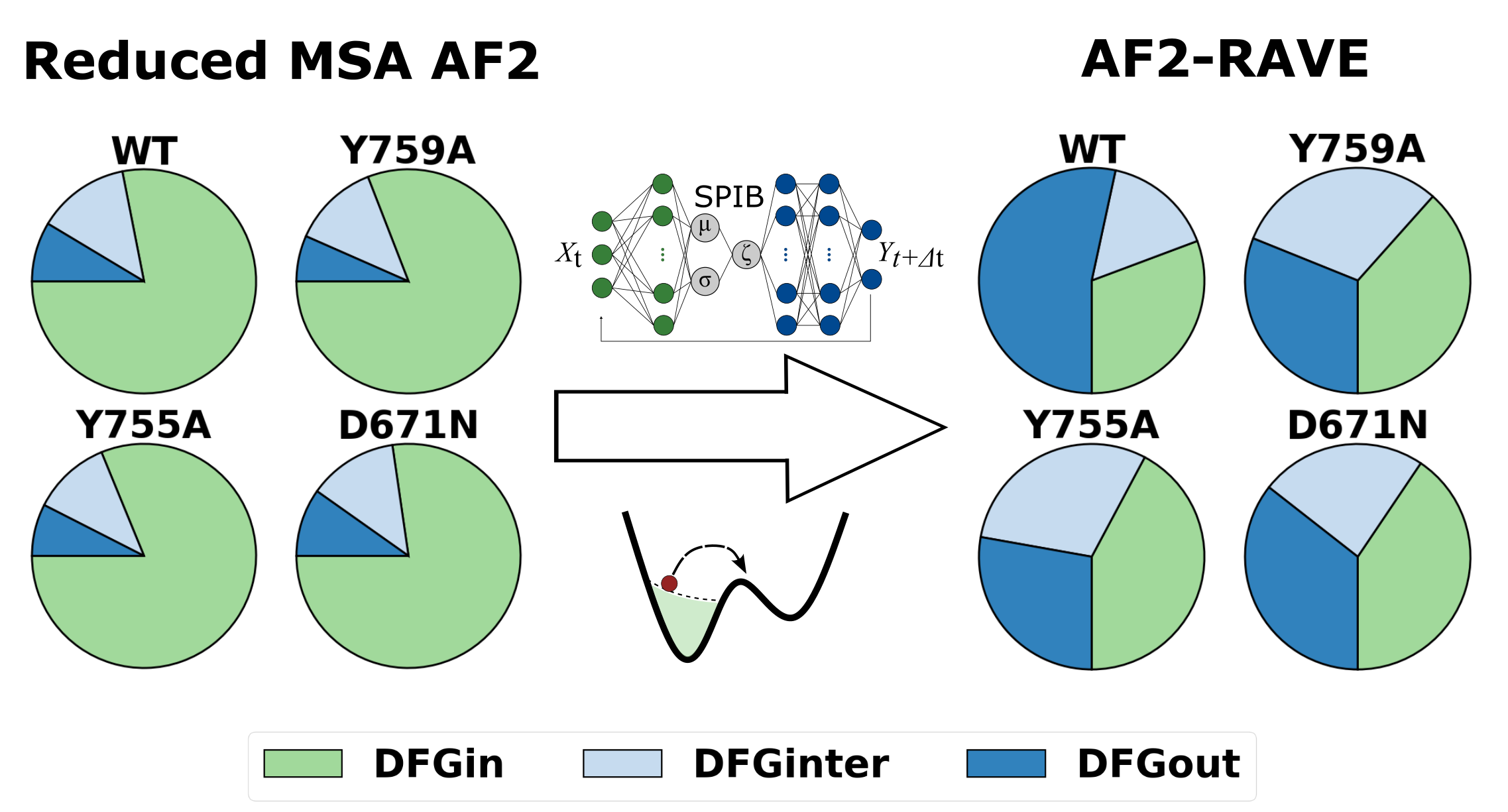}

\end{tocentry}

\begin{abstract}
Kinases compose one of the largest fractions of the human proteome, and their misfunction is implicated in many diseases, in particular cancers. The ubiquitousness and structural similarities of kinases makes specific and effective drug design difficult. In particular, conformational variability due to the evolutionarily conserved DFG motif adopting in and out conformations and the relative stabilities thereof are key in structure-based drug design for ATP competitive drugs. These relative conformational stabilities are extremely sensitive to small changes in sequence, and provide an important problem for sampling method development. Since the invention of AlphaFold2, the world of structure-based drug design has noticably changed. In spite of it being limited to crystal-like structure prediction, several methods have also leveraged its underlying architecture to improve dynamics and enhanced sampling of conformational ensembles, including AlphaFold2-RAVE. Here, we extend AlphaFold2-RAVE and apply it to a set of kinases: the wild type DDR1 sequence and three mutants with single point mutations that are known to behave drastically differently. We show that AlphaFold2-RAVE is able to efficiently recover the changes in relative stability using transferable learnt order parameters and potentials, thereby supplementing AlphaFold2 as a tool for exploration of Boltzmann-weighted protein conformations.

\end{abstract}
\cite{Meller2023}

\section{Introduction}

The first step of a typical structure-based drug design pipeline\cite{SBDD_1} is target protein structure prediction\cite{SBDD_2}. Traditionally this has been done using experimental techniques like x-ray crystallography, NMR spectroscopy or computational techniques like homology modeling \cite{NMR_prot,XRD_prot_rev,Homology_model}. These are all either time consuming, have limited accuracy or require adequate prior knowledge. With AlphaFold2\cite{Jumper}(AF2), we saw a paradigm shift in protein structure prediction. However, protein function is not solely dependent on a single native like structure, rather it is only properly understood or characterized by the protein's structural ensemble, including potentially several metastable conformations. Moreover, it is not sufficient to have a sense of conformational diversity alone, as  relative thermodynamic stabilities of protein conformations can be key in understanding activity, effects of mutation, and differences in behaviors of closely related proteins. In the short time since the development of AlphaFold2 (AF2), several publications have discovered ways to bridge the gaps between conformational dynamics variability and AF2 predictions. Many of these leverage AF2's internal architecture, and range over a spectrum between needing substantial input from physics-based simulation engines to not needing any physics at all. A common approach is to exploit the input featurisation of multiple sequence alignment to introduce stochastictity and deviations from native structure to the AF2 prediction\cite{AF2eLife.75751}. This includes our work\cite{VaniAA} combining AF2 and the machine learning-based enhanced sampling method Reweighted Autoencoded Variational Bayes for Enhanced Sampling (RAVE)\cite{Wang2019nc} into a combined protocol which we call AF2-RAVE to go from sequence to conformations ranked as per their thermodynamic or Boltzmann weights.

A well explored way to study the conformational diversity of a biomolecule is using the computational method of molecular dynamics (MD), i.e. by parametrizing intra- and inter-molecular forces with a force field and integrating newtons equations of motion\cite{Zwier_2010,Karplus_1990,Klepeis_2009}. However there are two key challenges in MD. First is the difficulty in sampling biologically relevant timescales. Since the integration of the equations of motion are limited by or fastest degree of motion which is at a femtosecond time scale, seeing changes of interest which can be at timescales of nanoseconds to hours is often prohibitively expensive and intractable with our current computational capacities. This has given rise to a large body of work in enhanced sampling algorithms\cite{tiwary2016review,He_nin_2022,Dickson_2010,kleiman2023adaptive} for difficult to sample distributions. These algorithms essentially attempt to sample a modified distribution and then reweight observables to obtain the correct statistics. This includes a wide range of methods addressing different concerns in sampling, each with its own set of challenges to deal with and it's own limitations. Broadly these methods can be classified in at least two ways: those that attempt to change the underlying Hamiltonian of the system\cite{Bonomi2009,bussi2020using,Tiwary2015}, and those which aim to statistically bias trajectories by splitting and resampling them\cite{stringRoux2021,Dickson2014,Vani2022}.

The second, closely related to the first, challenge in MD is the so-called curse of dimensionality. Biomolecular systems usually are roughly in the range of $10^3$ to $10^7$ atoms, leading to an unmanageably large number of degrees of freedom. We do not aim to sample the entire configuration space of these systems. However, it is commonly true that most biomolecules have a small number of low lying degrees of freedom, or a low lying manifold, that completely describes transitions of interest, or can separate conformational differences of biological relevance\cite{PhysRevLett.101.208101,Manifold_debenedetti}. This underlying manifold is rather confusingly referred to by many names, some commonly used are: reaction coordinate, alluding to the fact that the manifold traverses transitions between metastable states; collective variable, as it is usually a function of multiple coordinates of the system; order parameter, as it is used to parametrize different metastable states. In this work, we will refer to these degrees of freedom as ``collective variables" (CVs) when using them as inputs or a basis set, and ``order parameter" (OP) when referring to the finally obtained variable that we will bias along. Solving this second challenge thus has implications for the first as well.

While some of the aforementioned enhanced sampling methods attempt a generalized increase in sampling across every degree of freedom for every atom, for instance by increasing temperature, most of them aim to increase sampling along a specific manifold in the configuration space of the molecule. Even for the more generalized enhanced methods, actually quantifying the increase in sampling for high dimensional systems is difficult. Conversely, for methods that sample along predetermined manifolds, it is often the case that the wrong choices result in incomplete or incorrect configuration space sampling. Prior to the advent of machine learning in chemistry these low-lying degrees of freedom were almost always chosen by careful inspection and prior biophysical knowledge\cite{Zwanzig1961,Kawasaki1973,Antoszewski2020,Hong2018,Casasnovas2017,Clark2016}. Since the popularization of manifold learning methods, the identification of collective variables using ML has been a large field of interest\cite{mehdi2023enhanced,No2020,Rydzewski2023}. However most methods in this field are still best suited to very small set of problems and each have their own particular limitations \cite{mehdi2023enhanced, He_nin_2022}, in particular the need for \textit{a priori} information remains a frequent bottleneck.

In this work, we use the AF2-RAVE\cite{VaniAA} protocol, but with some significant refinements that make it more efficient, statistically robust, and transferable to mutations, suggesting it might be transferable within families of closely related proteins. We demonstrate this protocol on a kinase and its mutants.

Protein kinases, and in particular the DFG-in to DFG-out transition have been extensively studied using MD along with several enhanced sampling methods\cite{Roux2015,Jiang2022}. This family of enzymes is one of the most important therapeutic targets for structure-based drug design, as they are ubiquitous in the human proteome\cite{Manning2002}. Their main role is to mediate cell signalling in a large range of biomolecular processes at the cell level, in particular replication, hence implicating them in a majority of cancers. While there already exist several highly effective medicinal molecules for cancer therapy that function by targeting and inhibiting kinases\cite{Bhullar2018}, one could argue that at best we have scratched the surface in terms of kinase based therapeutics\cite{Kin_DD_21}.

In their active state, protein kinases catalyze the phosphorylation of substrate proteins through the transfer of the $\gamma$-phosphate group from adenosine triphosphate (ATP) or guanosine triphosphate (GTP). Often, the substrate protein is another in a cascade of kinases required for cell signalling\cite{Kin_pathway, DDR1_signal}. While there are over 500 kinases in the human kinome, making them a challenging class to study, the protein kinome universally has some highly conserved structural motifs with a structurally well characterized active state. One key motif for this characterization is the Asp-Phe-Gly (DFG) motif in the activation loop. This motif has two structural conformers, one with the Asp pointing into the loop, the DFG-in or active conformation, and one with it pointing out into the solvent, the DFG-out or inactive conformation. The ATP binding, and hence phosphorylation catalysis can only occur in the DFG-in conformation. Most drugs targeting kinases are ``ATP competitive", i.e. they bind to prohibit ATP binding. These ATP-competitive drugs themselves are classified mainly in two types, either binding to the active site in DFG-in conformation, hence inhibiting the binding of ATP, or binding to the DFG-out conformation and hence stabilizing the inactive state. However, the presence of kinases in many essential cellular functions and their homologous nature makes specificity and efficacy particularly hard to achieve. Often, we are interested in drugs that bind preferentially to specific kinases \textit{without} affecting other kinases. Given this, the characterization of diverse inactive states is of particular importance, both in terms of a structural understanding of these states, as well as knowledge of the thermodynamic stabilities relative to active state. For instance, given a target kinase, finding a uniquely stable inactive state with a novel binding site could lead to a more specific, less promiscuous (and hence toxic) drug. However, the most robust way to do this computationally, obtaining a MD trajectory that traverses the space of active and inactive states multiple times, is impossible. Additionally, the transition from DFG-in to DFG-out is highly non-local and a concerted combination of several long-range and large-scale motions, so that characterizing it and studying it even with the aid of enhanced sampling is difficult.

Another important covariate to study kinase conformational ensembles is that of point mutations, which are often the cause of incorrect signalling leading to pathology. One key reason for this is that the balance of probability between active and inactive conformations is delicate and often flipped on changing single residues, as shown by the system we have chosen in this work, DDR1.

Our AF2-RAVE based approach for solving this problem involves combining structural ensembles obtained from AF2 with a machine learning algorithm to learn order parameters for biasing. To demonstrate this, we use the discoidin domain receptor tyrosine kinase 1 (DDR1), which in wild-type is more stable in an inactive DFG-out conformation. However, in several single site mutations, specifically D671N, Y755A and Y759A, the relative DFG conformational stabilities are flipped. One of the reasons we choose this set of systems is that a recent paper\cite{Hanson2019} provides extremely detailed and valuable work on their DFG stabilities, with atypically long unbiased MD trajectories. We show that our results agree with theirs qualitatively, in that we predict the flipping of DFG conformation preference on mutation. While we sacrifice some accuracy qualitatively, our method is faster by roughly 2-3 orders of magnitude (exact simulation lengths described in Results), and has the potential of being reused without relearning a lot of the information.

We will begin by discussing the methods used within the protocol and outlining AF2-RAVE. Next, we discuss some molecular biology background for kinases, in particular those which may be important to the DFG-in to DFG-out transition. Finally, we will describe our protocol and the results we obtained.

\section{Methods}

In the section we will first describe the methods that compose our protocol: \hyperref[sec:AF2MSA]{(i)} AlphaFold2 and MSA depth modification, \hyperref[sec:MetaD]{(ii)} Metadynamics, and \hyperref[sec:SPIB]{(iii)}State predictive information bottleneck (SPIB\cite{Dedi}, the most recent variant of RAVE\cite{Wang2019nc}). Finally we will list important parameters to note for our MD simulations in \hyperref[sec:simulation]{(iv)} Simulation Details. Additionally, Fig. 1 shows a high level flowchart form of the method.

\begin{figure*}[t!]
    \centering
    \includegraphics{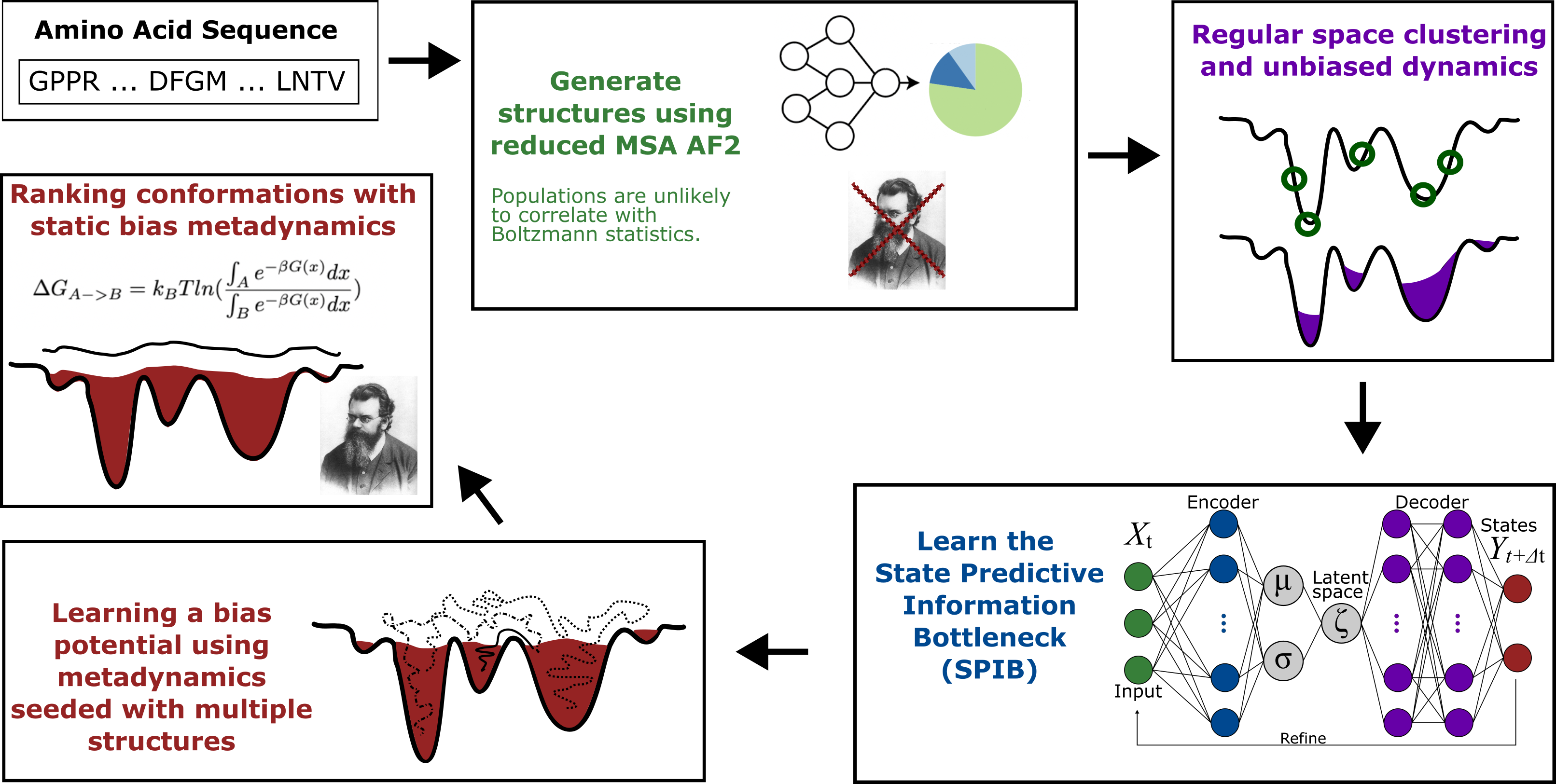}
    \caption{A high level schematic of the method, showing: (i) a typical input sequence, (ii) AF2 generated seed structures, (iii) regular space clustering and unbiased runs, (iv) SPIB to suggest OPs, (v) metadynamics runs.}
    \label{fig:scheme}
\end{figure*}

\subsection{AlphaFold2 and MSA depth modification}
\label{sec:AF2MSA}
The search for a computational model to predict crystal structures or other native-like structures for proteins has been a central part of computational molecular biology. When AF2 was introduced in 2020 it was unprecedented in its speed and accuracy\cite{Jumper}. The internal architecture of AlphaFold2 uses three primary components: the alignment of multiple evolutionarily related sequences, an attention-based neural network, or transformer, and a black hole initialized  attention based graph neural network structure module. The model is trained on the entire RCSB database protein database of experimentally derived structures. While transformative to the field, the model does not quite solve the protein folding problem, as proteins \textit{in vivo} are not defined by a single structure but by the structural ensemble.

The multiple sequence alignment (MSA) form of the input has been found to be a convenient point of input to introduce stochasticity. The simplest way to do this, which we employ in AF2-RAVE is to decrease the depth of the MSA input into both channels of the model, and then run the model repeatedly with randomly chosen subsets of the full MSA. In some sense this process is a way to withhold data from the model to produce ostensibly incorrect outputs. However, AF2 has also been found to perform consistently badly in situations that deviate from the norm of their evolutionarily related sequences, which is likely due to the MSA form of input featurization leading to significant bias. From this perspective, the above described protocol should lead to some random sampling of the correct structure in these special cases. Moreover, in cases where the protein family has multiple metastable structures, with the structure of native stability being different in different members of the family, this protocol could conceivably provide hints or ``breadcrumbs" for the entire conformational space of interest.

A recent study\cite{Roney2022} has proposed that AF2 has indeed learned an energy-surface for protein folding in its transformer weights. They propose and provide significant evidence for the idea that the MSA pair representation matrix simply initializes close to the correct minimum while the transformer architecture performs an optimization step on this energy surface. This suggests that our MSA reduction protocol simply initializes the transformer closer to a different local minimum, possibly a non-native for the learned energy surface. This might be a biologically native-like structure from the previously described special cases, or a biologically relevant metastable structure.

In spite of this, this modified version of AF2 still leads to some highly unphysical structures, as we illustrated previously\cite{VaniAA}. Worse still, the structures obtained, including those that are metastable are not in any physically reasonable probability distribution. Nor is there an obvious way to directly obtain a distribution or free energy surface from them that could account for both enthalpy and entropy. Some direct notion of physics and thermodynamics is still required for this information to be usable.

\subsection{Metadynamics} \label{sec:MetaD}

Hamiltonian-based enhanced sampling algorithms rely on the approach of editing the underlying energetics of dynamical systems. For instance umbrella sampling adds harmonic restraints along successive points of conformation of space in replicate simulations and re-combines them to reproduce the original energy surface. One of the most powerful of this family of methods to explore complex energy landscapes with high barriers is metadynamics\cite{Tiwary2015}.

For a predetermined order parameter(s), metadynamics aims to learn a biased potential that is the negative of the true free energy. To achieve this, a history dependent potential is added to the Hamiltonian of the dynamics. This potential is updated periodically by adding Gaussian functions to the bias function centered at current values of the order parameters.

This potential acts as a driving force, pushing the system away from visited regions, forcing it to explore new areas of the configuration space. The specific version we employ is well-tempered metadynamics, wherein the height of the Gaussians is modulated with a time dependence to mimic a high temperature simulation and to prevent the bias potential from growing indefinitely. In this case the bias potential can be shown to converge to the true underlying free energy modulo a multiplicative constant\cite{Dama2014}.

While clever and asymptotically accurate estimates for time dependent bias rewinding exist for dynamic biases\cite{Tiwary2015}, they are subject to normalization errors and some strong assumptions. Additionally in spite of the well-tempering of this method, often its unbounded nature can result in sampling regions of configuration space that we are not interested in or are sufficiently rare enough to be irrelevant. In the study we use metadynamics to learn a potential that we then freeze and use as a static Hamiltonian bias. By controlling the region explored by the initial metadynamics through a stop condition we circumvent learning and unbounded potential bias, and compute our final statistical estimates with fewer errors. Next, we initialize independent walkers using the same static bias from both DFG-in and DFG-out structures, and run them till we see transitions followed by a stable trajectory in the basin that they were not initialized in. In general, since metadynamics relies on dynamically pushing the simulation to undiscovered regions, and has historically been considered difficult to learn an effective static bias, and hence is not a common practice.\cite{LamimRibeiro2018,Dama2014} Every single independently launched trajectory visit the DFG-in basin when launched from DFG-out, and the DFG-out basin when launched from DFG-in. We find this to be a computationally more efficient protocol for getting overlap in explored configuration space. We find that the same static bias can also occasionally lead to back-and-forth transitions in the same trajectory, but the approach used in this work is computationally much more attractive.

It is important to note however that metadynamics suffers from the common enhanced sampling method limitation that it requires an \textit{a priori} notion of the approximate reaction coordinate or underlying low dimensional manifolds to use as order parameters for sampling. A now common approach that we also employ in this work is to use a machine learning method to learn OPs for sampling, specifically, here we use the approach of a time lagged state predictive autoencoder, described below. Previous work has shown its suitability to learn metadynamics OPs for a range of systems such as conformational changes, membrane permeation\cite{Mehdi2022AiB}, as well as for using in this AF2-seeded approach\cite{VaniAA}. This represents a crucial step towards making enhanced sampling methods usable on novel systems with limited \textit{a priori} biophysical understanding.

\subsection{State predictive information bottleneck} \label{sec:SPIB}

To solve the problem of an unknown underlying manifold required for biasing that captures relevant slow degrees of freedom, we use a method based originally on the reweighted autoencoder for variational Bayes algorithm (RAVE)\cite{Wang2019nc}. We use its more updated form, the state predictive information bottleneck\cite{Dedi}.

In general a variational auto encoder (VAE) is a neural network framework that attempts to learn a low dimensional probabilistic function (the encoder) of the input and a function that is then able to reproduce the input (the decoder). This underlying low dimensional function is the information bottleneck, i.e. it minimizes input information while maximizing its ability to obtain the output. Here, since we aim to study a dynamical trajectory, we modify the basic VAE to incorporate a past future information bottleneck. Given a frame of the trajectory as an input, instead of reproducing this input, we reproduce a trajectory frame at a later time stamp, i.e. with some time lag. Additionally, we note that since proteins have several degrees of freedom that move at different time scales, for a specific time lag we are not attempting to reproduce the trajectory at every coordinate. Since we do not know which degrees of freedom correspond with which time scale, we instead choose for the output a notion of states, represented by one hot encoded vectors. These states are iteratively learnt between epochs of training the neural network. This protocol has been shown to be effective in several complex systems\cite{Mehdi2022AiB,Beyerle_ent_ener,Zou2023}.

Since we aim to make our protocol generalizable, we start with a large set of input CVs. However, biasing using metadynamics on a function of a large set of CVs is hard to control and not always statistically stable. To alleviate this to some extent, we adopt a basis CV refinement step as a stand-in for regularization in our OP learning protocol, wherein we run SPIB three times, each time discarding features with weights lower than 0.25 of the maximum weight.

\subsection{Simulation details} \label{sec:simulation}

\begin{figure*}[t!]
    \centering
    \includegraphics{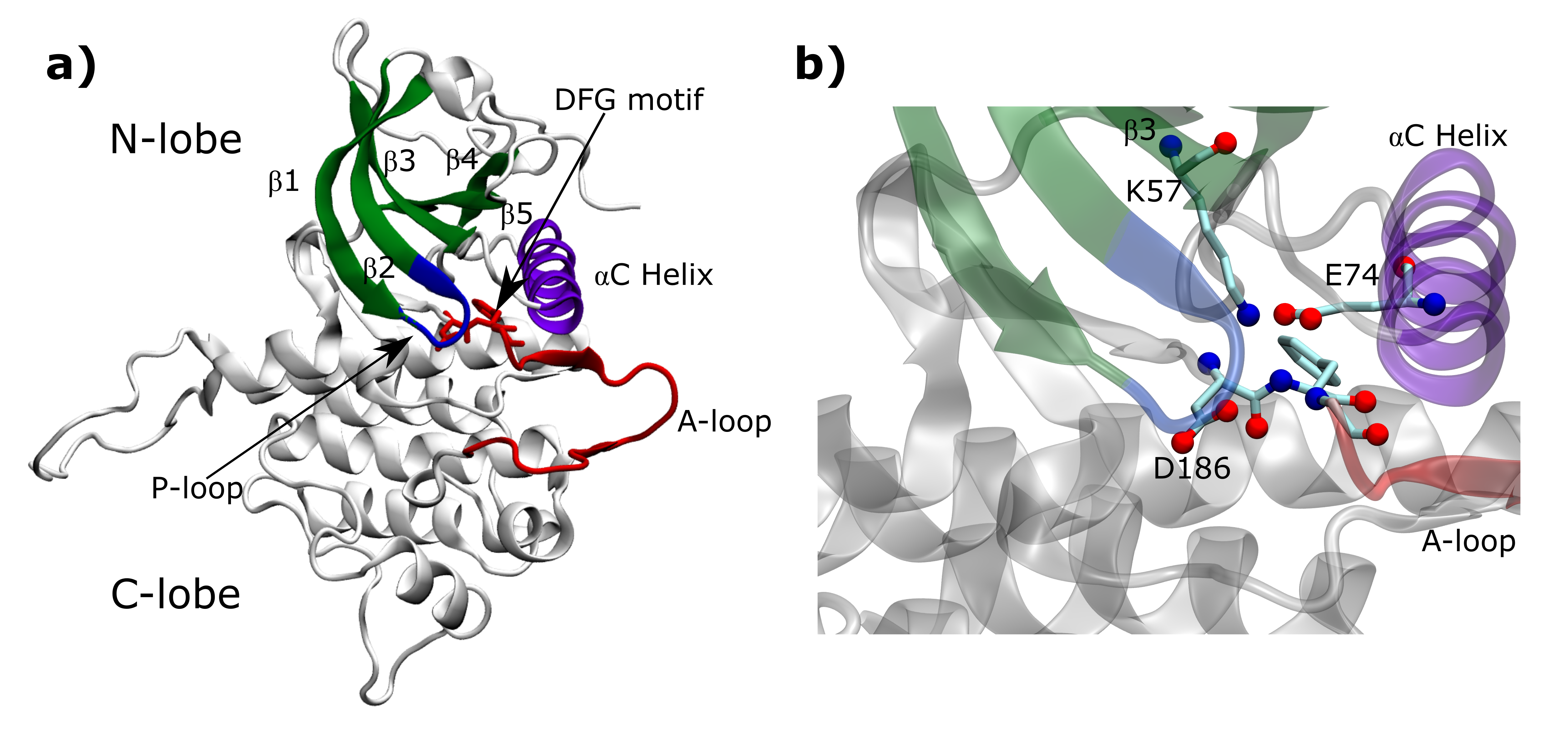}
    \caption{a) The structural anatomy of the DDR1 kinase molecule showing the activation loop (red), Gly rich P-loop(blue), $\alpha$C helix (purple) and the characteristic N lobe $\beta$ sheets (green). These motifs are relevant to DFG-in to DFG-out transition. b) The conserved salt bridge between K57 in $\beta_3$ and E74 in $\alpha$C helix that is crucial in ligand dissociation and for basic kinase functioning\cite{Tiwary2017}.}
    \label{fig:Chel}
\end{figure*}

The protein is represented by the AMBER03 force field\cite{amberff03}. The simulations are performed at 300 K with the BAOAB integrator\cite{Leimkuhler2013} in OpenMM\cite{Openmm_1}; LINCS is used to constrain the lengths of bonds to hydrogen atoms\cite{hess1997lincs}; Particle Mesh Ewald is used to calculate electrostatics\cite{PME}; the step size was 2 fs. The systems are solvated with TIP3P water models and equilibrated under NVT and NPT for 200ps and 300ps respectively. To prevent melting during biasing, we also restrain the conserved $\alpha$C-helix of the N-lobe. Large scale motion for this motif is essential, both to see a transition and for drug unbinding pathways\cite{Tiwary2017,Shekhar2022}. However, biasing CVs that include distances from this helix often leads to irreversible disordering, and we find that applying torsional restraints on residues 65 to 81 is sufficient to allow for smooth upward motion. In Figure \ref{fig:Chel}, we show the $\alpha$C-helix and its position with respect to the DFG loop.

\section{Results and Discussion}

Our overall protocol comprises the methods described in the previous section. We start by using reduced-MSA AlphaFold2 to generate a diverse set of initial conformations, and cluster them using regular space clustering. We choose this method of clustering because the reduced-MSA AF2 outputs tend to be quite sparse in regions of interest, and we want to prioritize the tails of the distribution over highly sampled regions. The centers of these structures are then used to seed unbiased MD trajectories, which are our input trajectories for RAVE. This learns a 2-dimensional order parameter expressed as an information bottleneck. Then, we run well-tempered metadynamics biasing along this 2-dimensional order parameter. Finally we use the metadynamics learnt bias for the wild type protein as a static bias to sample distributions for the wild type and mutant structures sequences.

We will discuss the results in two sections. First, we discuss the modified AF2 outputs for all four sequences, and the process of learning a biasing potential, and next, we discuss results from biased dynamics for DDR1 WT and mutants. 

\begin{figure*}
    \centering
    \includegraphics[width=\textwidth]{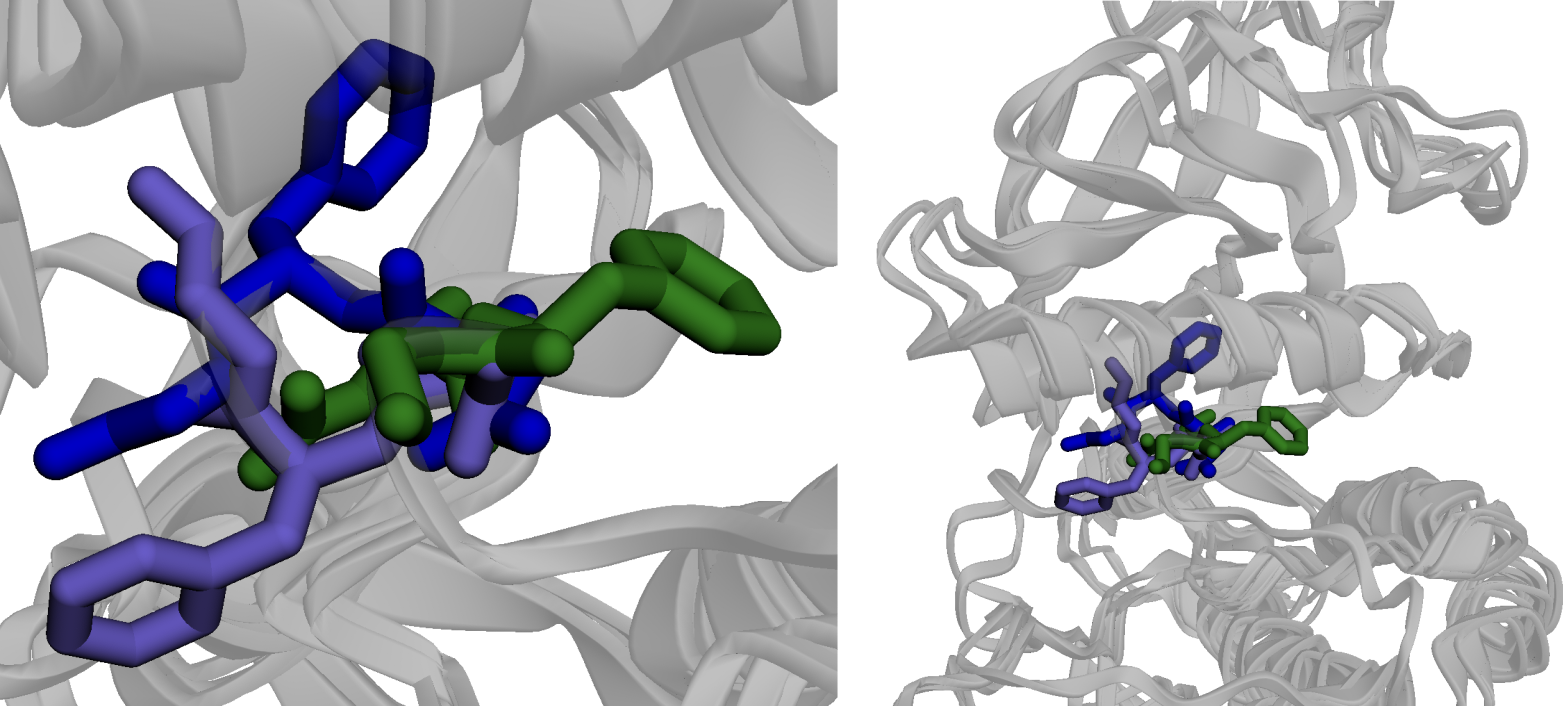}
    \caption{Representative structures from reduced MSA AF2 for the DFG-in (purple), DFG-inter (blue) and DFG-out (green) shown in two views, superimposed on the same structure.}
    \label{fig:dfgexplain}
\end{figure*}

When we refer to the kinase DFG-in, -out, and -inter structures, we use the Dunbrack method \cite{Modi2021} for classification using distance cutoffs. Representative structures for these states are shown in Fig. \ref{fig:dfgexplain}. When referring to dimensions or OPs that are learnt through RAVE, we label them as information bottlenecks (IB) and assign variables $\sigma$.

\subsection{Learning a bias potential from AF2-RAVE on WT DDR1}
\label{sec:learn}

Our first step is to generate structures using the reduced-MSA version of AF2. We generate 1280 structures for each kinase: 640 for MSAs of depth 16 and 32, with 128 random seeds generating 5 structures each. AlphaFold2 even with the reduced MSA approach is simply unable to distinguish between the conformational diversity expected for these 4 sequences, giving effectively identical results for all. This can be seen from Fig. \ref{fig:AF2DFG}a) where we show the populations of active, inactive, and the known intermediate or transition state ``DFG-up" or ``DFG-inter" state after filtering out obviously unphysical structures (e.g. with broken bonds). These structures are classified using Dunbrack classification described below. We note that while we see significant DFG-out population, even in the wild type which is known to have a higher inactive state stability, AF2 predicts the more commonly found DFG-in structure. These are in disagreement with population densities implied by previous long MD simulations of the same kinase, in spite of thoroughly searching MSA hyperparameters to force increased structural diversity.\cite{Hanson2019} The transition state is currently commonly called ``inter"\cite{Modi2021}, as it has been consistently found to be necessary structure to see the DFG-in to DFG-out trajectory. Previously, it was referred to as ``DFG-up", for the upward pointing position of the Asp residue sidechain in the traditional structural view (with the N-lobe above the C-lobe), while the previously named ``DFG-down" position is referred to as unassigned, as it is a high energy, physically unlikely structure. The fact that we see the transition state from reduced MSA AlphaFold2 is extremely significant and useful, as we have previously attempted to use RAVE simply with crystal structures of DFG-in and -out structures and failed, as simulations tend to push towards and then get stuck in the ``DFG-down" configuration. We show further analysis of the AF2 reduced-MSA outputs in the SI.

Next, we use SPIB on these structures to propose possible OPs for biasing. The active-inactive conformational transition is a highly delocalized one, and requires an understanding of several intramolecular interactions. In particular, the $\alpha$ C-helix forms a conserved salt bridge with the $\beta3$ strand that plays a crucial role in the molecule's transition \cite{Tiwary2017, Modi2019}. Further, globally, the opening and closing of the two lobes (N and C) define the active-inactive transition\cite{Dyn_Pers}. From a dynamics perspective, the prime motifs involved in this transition include the Activation loop (A-Loop), P-Loop and $\alpha$C-helix, which mainly belong to the N-lobe (Fig. S1). To this end, we include a number of other CVs, described below, including the distances used for the DFG classification above. These CVs roughly correspond to those used in several previous studies focusing on different parts of the kinase molecular structure\cite{Modi2021,Narayan2020,Saladino2012,Meng2018}. In Table S1, we list all the residues involved in distances that we consider, indicating our acronym, the conserved residue (if applicable), and the resid for DDR1 (numbered from 1), and a description of the motif it belongs to. In Table S2, we list all the distances used as initial inputs for SPIB, and indicate them visually in Figure S1.

In this nomenclature, the Dunbrack distances, which we use to project our final potentials of mean force (PMFs), are sbridgeK CB - ChelE CB  and sbridgeK CB - DFGAsp CB. Referring to these as $d_1$ and $d_2$, the structures are classified as: DFG-in if $d_1<11$ and $d_2>14$, DFG-out if $d_1>11$ and $d_2<14$, DFG-inter if $d_1<11$ and $d_2<11$, and unassigned if $d_1>11$ and $d_2>11$.

In Fig. S2 we show the distributions of AF2 outputs for all four sequences of DDR1 on the entire set of input CVs and find that they all look quite similar, in spite of known differences in these mutants. This is interesting, considering we expect AF2 to usually perform more confidently in evolutionarily faithful sequences. However, in this case, we suspect that the fact that DDR1 is unusual in its DFG-out stability contributes to the easy access it has to conformational diversity. Nevertheless, our AF2 outputs still predict higher stability for the DFG-in structures universally for the DDR1s.

To learn a biasing potential, we run multiple metadynamics trajectories in parallel that share bias potentials. This is done only for the wild-type sequence and the same bias is then used for further calculations for all sequences. These initial structures are chosen as in the original AF2-RAVE paper\cite{VaniAA}, as follows: first, we run AF2 using ColabFold with a manually set MSA depth of 8 and 16. Next, we run regular space clustering with the minimum distance parameter of $9$ on standardized CV values, using the set of CVs described above.

We use the following stopping conditions for the initial metadynamics runs: (1) If the walker started in DFG-in(out) structures, it must reach the DFG-out(in) structure, and (2) If the walker was not initially in one of the two main metastable states, it must reach one of them. We only stop if the transition is stable for 1 nanosecond during the biased simulation. These simulations are between 5ns and 20ns long.

In Fig. S4, we show the bias learnt for the process. In order to demonstrate the transferability and efficiency of our protocol, we only learn an IB and a bias potential using the wild type. This also provides a basis to propose learning universal biases for systems with transferable CVs potentially allowing for more efficient sampling for homologous families of proteins.

\subsection{Biased dynamics on DDR1 mutants}
\label{sec:bias}

\begin{figure*}[t!]
    \centering
    \includegraphics[width=\textwidth]{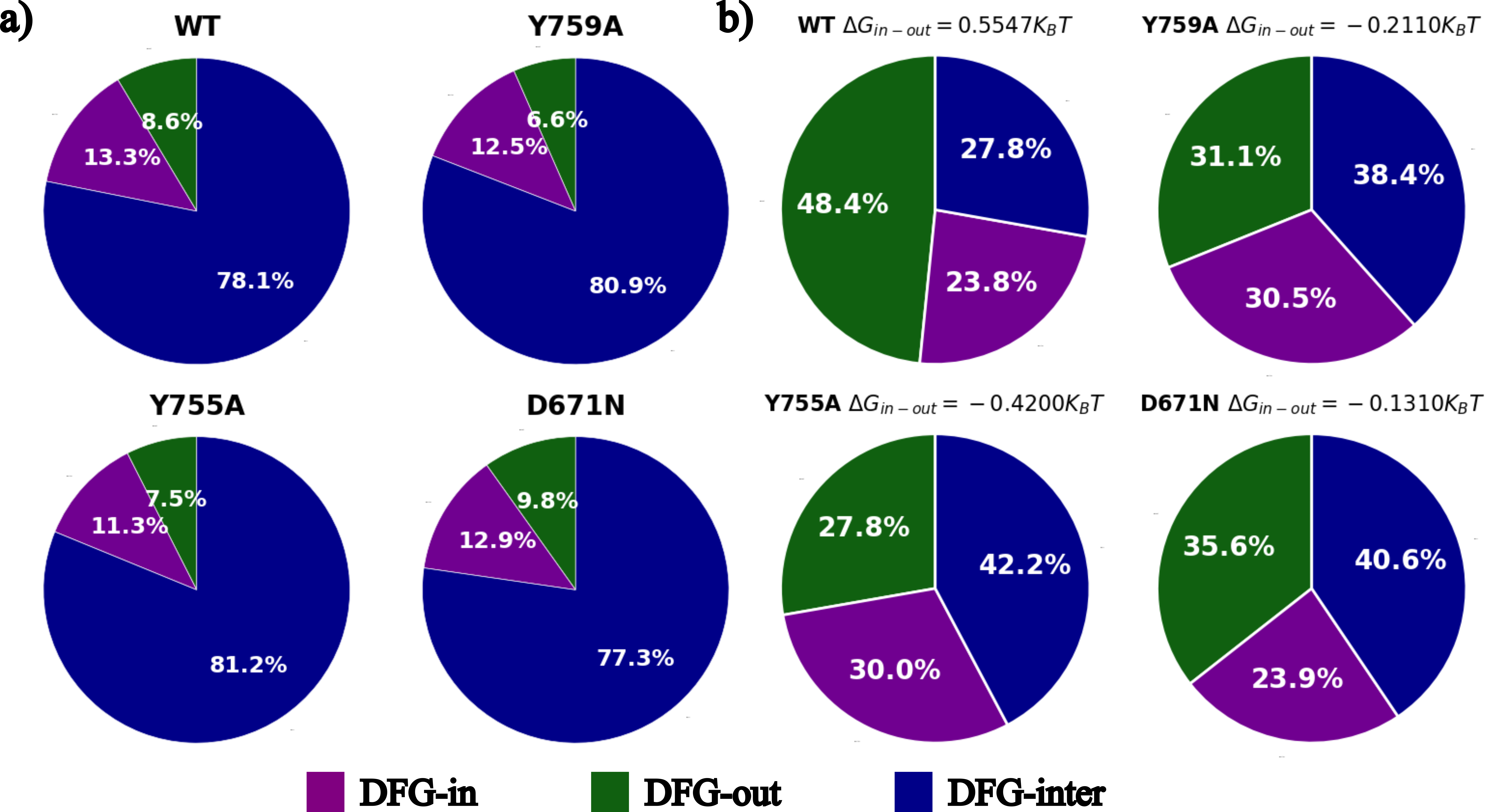}
    \caption{Populations of active, inactive, and the known transition state ``DFG-up" or ``DFG-inter" state for wild type and mutants D671N, Y755A, and Y759A, a) through reduced MSA AF2 (MSA lengths 8 and 16 combined) and b) using our AF2-RAVE protocol. We clearly see that AF2 by itself is unable to distinguish between wild type and mutants, and in particular for mutant gives us the wrong order of stability. On the other hand, AF2-RAVE is able to find the reversal in stability on point mutation, and give us more thermodynamically representative populations for these states, in excellent agreement with benchmark calculations performed in Hanson et al\cite{Hanson2019} using unbiased MD simulations.}
    \label{fig:AF2DFG}
\end{figure*}

In this study, our main result is the accurately predicted relative stabilities of the DFG-in and DFG-out structures for the wild type and mutants for the kinase DDR1. In Fig. S5, we show predicted PMFs for these structures. We compute DFG-in versus DFG-out relative thermodynamic stability by integrating probabilities over Dunbrack definitions for kinase structural states and calculating the $\Delta G$ between these states: (i) WT: 0.5 $K_BT$, (ii) D671N: -0.2 $K_BT$, (iii) Y755A: -0.42 $K_BT$, (iv) Y759A: -0.13 $K_BT$. Each of these were computed with 5 trajectories each, and had standard deviations of (i) WT: 0.23 $K_BT$, (ii) D671N: 0.12 $K_BT$, (iii) Y755A: 0.29 $K_BT$, (iv) Y759A: 0.11 $K_BT$. This flipping in Boltzmann ranking of active and inactive is in concurrence with previous findings\cite{Hanson2019}, which were obtained using long unbiased MD simulations. We can also integrate over our reweighted data to obtain the Dunbrack populations to compare with those from AF2, shown in Fig. \ref{fig:AF2DFG}.

\begin{figure*}[t!]
    \centering
    \includegraphics[width=\textwidth]{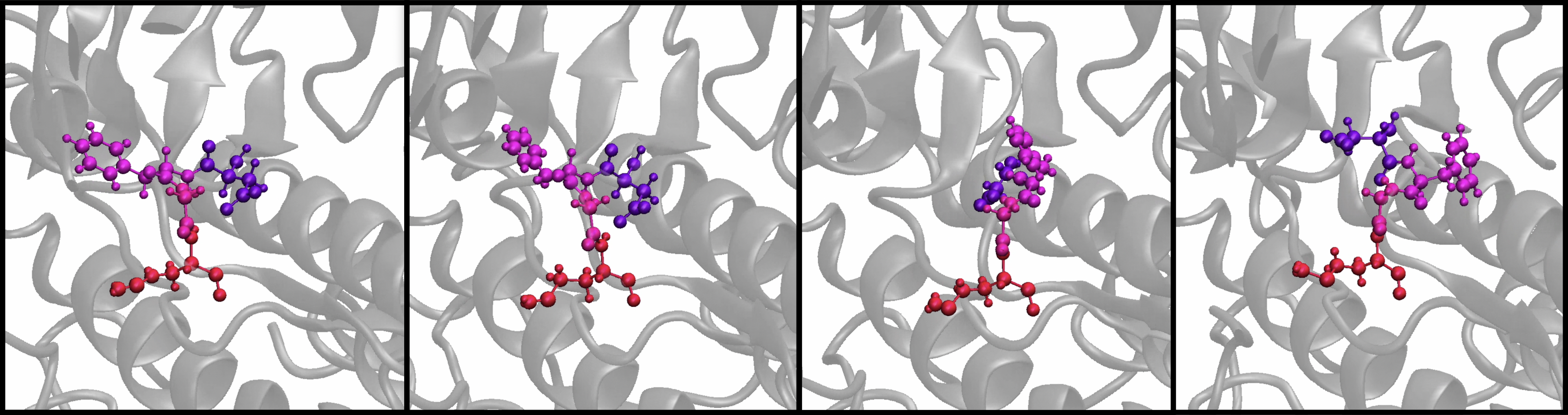}
    \caption{An example trajectory for the DFG-out to DFG-in transition obtained using AF2-RAVE. See Code Availability for details of an example video showing the transition.}
    \label{fig:traj}
\end{figure*}

In Fig. \ref{fig:traj} we show snapshots of one example of our sampled trajectories. We also note some salient details of the transition we study. We find that in the process of the transition, the breakage and formation of the salt bridge is clear, and the large scale upward motion of the Chelix is absolutely essential for sampling dynamics, as noted previously.

\section{Conclusion}

In this work, we have extended our previous protocol AF2-RAVE to obtain Boltzmann-ranked conformational diversity in protein kinases, specifically from the perspective of the pharmaceutically relevant and evolutionarily conserved DFG loop. We have previously shown that the protocol is able to capture a versatile range of transitions: rotameric metastability, large scale helical motions, and partial disordering\cite{VaniAA}. Here, we study a the DFG loop conformational change that is of utmost therapeutic importance. Moreover, we choose DDR1 since there exists an unusually thorough sampling with unbiased trajectories by Hansen et al.\cite{Hanson2019} which provides more robust comparison than enhanced sampling-based work on large biomolecules is usually afforded. With AF2-RAVE we obtain the same conformational ranking and similar thermodynamic stabilities as found by Hanson et al.\cite{Hanson2019} for the DFG-in versus DFG-out conformations of DDR1. However the total MD simulation time in our work is around 2-3 orders of magnitude shorter.

It is important to note that our trajectories rely on collective variables that allow us to sample the DFG-inter transition state. Without AF2 structures to seed our initial unbiased trajectories, the collective variable learnt usually leads to a DFG-down structure resulting in an unsuccessful transition trajectory. Another significant factor of our work is the use of a static learnt bias, which is usually considered difficult and avoided. However, we find that the restriction to transitions of importance is necessary for this measurement in terms of both speed and replication.

This entire protocol can be repeated for other kinases in a few different ways. The first is to learn a bias using a wild-type and then sample mutants that are known to be pathological or disease causing due to changes in activity. The second is to learn a bias using a single kinase and sample other closely related kinases using the same potential. Finally, we hope that eventually a generalized set of collective variables and biases can be learnt that could sample across the human kinome. This protocol can be used to sample novel states that can then be used as relatively high-throughput inputs for cryptic pocket prediction algorithms\cite{GrASP}, and we demonstrate some examples in the SI.

\textbf{Acknowledgments}

P.T. and B.V. were supported by the National Institute of General Medical Sciences of the National Institutes of Health under Award Number R35GM142719. The content is solely the responsibility of the authors and does not represent the official views of the National Institutes of Health. A.A. was supported by NCI-UMD Partnership for Integrative Cancer Research. We are grateful to NSF ACCESS Bridges2 (project CHE180053) and University of Maryland Zaratan High-Performance Computing cluster for enabling the work performed here. 
We thank Drs. Eric Beyerle, Xinyu Gu, Zack Smith, and Dedi Wang for critical reading of the manuscript, and Drs. Mrinal Shekhar, Shashank Pant, Zack Smith and Dedi Wang for helpful discussions regarding kinases.
\newline 

\textbf{Notes}

The authors declare the following competing financial interest(s): P.T. is a consultant to Schrodinger, Inc. and is on their Scientific Advisory Board. 

\begin{suppinfo}

Detailed description of methods used and further analysis of systems and sampling can be found in the supplement.

\section{Code Availability}

 The code to run AF2-RAVE in a seamless manner is available at \url{https://github.com/tiwarylab/alphafold2rave}. This can be run on Google Colab using GPUs. Using Colab Pro is advised. Codes, parameters, and bias files used to specifically run the simulations from this protocol can be found at \url{https://github.com/tiwarylab/kinase_Aloop}. These will currently only work for the sequences used in this paper, but most files are easily adaptable for use in other kinases with some specific changes, which are marked. An example video is also in the folder, and while full trajectories are too large to upload, they can be made available on request.

\section{Data Availability}
All data associated with this work is available through \url{https://github.com/tiwarylab/kinase_Aloop}.

\end{suppinfo}
\providecommand{\latin}[1]{#1}
\makeatletter
\providecommand{\doi}
  {\begingroup\let\do\@makeother\dospecials
  \catcode`\{=1 \catcode`\}=2 \doi@aux}
\providecommand{\doi@aux}[1]{\endgroup\texttt{#1}}
\makeatother
\providecommand*\mcitethebibliography{\thebibliography}
\csname @ifundefined\endcsname{endmcitethebibliography}
  {\let\endmcitethebibliography\endthebibliography}{}


\begin{mcitethebibliography}{64}
\providecommand*\natexlab[1]{#1}
\providecommand*\mciteSetBstSublistMode[1]{}
\providecommand*\mciteSetBstMaxWidthForm[2]{}
\providecommand*\mciteBstWouldAddEndPuncttrue
  {\def\EndOfBibitem{\unskip.}}
\providecommand*\mciteBstWouldAddEndPunctfalse
  {\let\EndOfBibitem\relax}
\providecommand*\mciteSetBstMidEndSepPunct[3]{}
\providecommand*\mciteSetBstSublistLabelBeginEnd[3]{}
\providecommand*\EndOfBibitem{}
\mciteSetBstSublistMode{f}
\mciteSetBstMaxWidthForm{subitem}{(\alph{mcitesubitemcount})}
\mciteSetBstSublistLabelBeginEnd
  {\mcitemaxwidthsubitemform\space}
  {\relax}
  {\relax}

\bibitem[Meller \latin{et~al.}(2023)Meller, Bhakat, Solieva, and
  Bowman]{Meller2023}
Meller,~A.; Bhakat,~S.; Solieva,~S.; Bowman,~G.~R. \emph{Journal of Chemical
  Theory and Computation} \textbf{2023}, \emph{19}, 4355--4363\relax
\mciteBstWouldAddEndPuncttrue
\mciteSetBstMidEndSepPunct{\mcitedefaultmidpunct}
{\mcitedefaultendpunct}{\mcitedefaultseppunct}\relax
\EndOfBibitem
\bibitem[Ferreira \latin{et~al.}(2015)Ferreira, dos Santos, Oliva, and
  Andricopulo]{SBDD_1}
Ferreira,~L.; dos Santos,~R.; Oliva,~G.; Andricopulo,~A. \emph{Molecules}
  \textbf{2015}, \emph{20}, 13384--13421\relax
\mciteBstWouldAddEndPuncttrue
\mciteSetBstMidEndSepPunct{\mcitedefaultmidpunct}
{\mcitedefaultendpunct}{\mcitedefaultseppunct}\relax
\EndOfBibitem
\bibitem[Batool \latin{et~al.}(2019)Batool, Ahmad, and Choi]{SBDD_2}
Batool,~M.; Ahmad,~B.; Choi,~S. \emph{International Journal of Molecular
  Sciences} \textbf{2019}, \emph{20}, 2783\relax
\mciteBstWouldAddEndPuncttrue
\mciteSetBstMidEndSepPunct{\mcitedefaultmidpunct}
{\mcitedefaultendpunct}{\mcitedefaultseppunct}\relax
\EndOfBibitem
\bibitem[W\"{u}thrich(1990)]{NMR_prot}
W\"{u}thrich,~K. \emph{Journal of Biological Chemistry} \textbf{1990},
  \emph{265}, 22059--22062\relax
\mciteBstWouldAddEndPuncttrue
\mciteSetBstMidEndSepPunct{\mcitedefaultmidpunct}
{\mcitedefaultendpunct}{\mcitedefaultseppunct}\relax
\EndOfBibitem
\bibitem[Shi(2014)]{XRD_prot_rev}
Shi,~Y. \emph{Cell} \textbf{2014}, \emph{159}, 995--1014\relax
\mciteBstWouldAddEndPuncttrue
\mciteSetBstMidEndSepPunct{\mcitedefaultmidpunct}
{\mcitedefaultendpunct}{\mcitedefaultseppunct}\relax
\EndOfBibitem
\bibitem[Cavasotto and Phatak(2009)Cavasotto, and Phatak]{Homology_model}
Cavasotto,~C.~N.; Phatak,~S.~S. \emph{Drug Discovery Today} \textbf{2009},
  \emph{14}, 676--683\relax
\mciteBstWouldAddEndPuncttrue
\mciteSetBstMidEndSepPunct{\mcitedefaultmidpunct}
{\mcitedefaultendpunct}{\mcitedefaultseppunct}\relax
\EndOfBibitem
\bibitem[Jumper \latin{et~al.}(2021)Jumper, Evans, Pritzel, Green, Figurnov,
  Ronneberger, Tunyasuvunakool, Bates, {\v{Z}}{\'i}dek, Potapenko, Bridgland,
  Meyer, Kohl, Ballard, Cowie, Romera-Paredes, Nikolov, Jain, Adler, Back,
  Petersen, Reiman, Clancy, Zielinski, Steinegger, Pacholska, Berghammer,
  Bodenstein, Silver, Vinyals, Senior, Kavukcuoglu, Kohli, and
  Hassabis]{Jumper}
Jumper,~J. \latin{et~al.}  \emph{Nature} \textbf{2021}, \emph{596},
  583--589\relax
\mciteBstWouldAddEndPuncttrue
\mciteSetBstMidEndSepPunct{\mcitedefaultmidpunct}
{\mcitedefaultendpunct}{\mcitedefaultseppunct}\relax
\EndOfBibitem
\bibitem[del Alamo \latin{et~al.}(2022)del Alamo, Sala, Mchaourab, and
  Meiler]{AF2eLife.75751}
del Alamo,~D.; Sala,~D.; Mchaourab,~H.~S.; Meiler,~J. \emph{eLife}
  \textbf{2022}, \emph{11}, e75751\relax
\mciteBstWouldAddEndPuncttrue
\mciteSetBstMidEndSepPunct{\mcitedefaultmidpunct}
{\mcitedefaultendpunct}{\mcitedefaultseppunct}\relax
\EndOfBibitem
\bibitem[Vani \latin{et~al.}(2023)Vani, Aranganathan, Wang, and Tiwary]{VaniAA}
Vani,~B.~P.; Aranganathan,~A.; Wang,~D.; Tiwary,~P. \emph{Journal of Chemical
  Theory and Computation} \textbf{2023}, \relax
\mciteBstWouldAddEndPunctfalse
\mciteSetBstMidEndSepPunct{\mcitedefaultmidpunct}
{}{\mcitedefaultseppunct}\relax
\EndOfBibitem
\bibitem[Wang \latin{et~al.}(2019)Wang, Ribeiro, and Tiwary]{Wang2019nc}
Wang,~Y.; Ribeiro,~J. M.~L.; Tiwary,~P. \emph{Nature Communications}
  \textbf{2019}, \emph{10}, 3573\relax
\mciteBstWouldAddEndPuncttrue
\mciteSetBstMidEndSepPunct{\mcitedefaultmidpunct}
{\mcitedefaultendpunct}{\mcitedefaultseppunct}\relax
\EndOfBibitem
\bibitem[Zwier and Chong(2010)Zwier, and Chong]{Zwier_2010}
Zwier,~M.~C.; Chong,~L.~T. \emph{Current Opinion in Pharmacology}
  \textbf{2010}, \emph{10}, 745--752\relax
\mciteBstWouldAddEndPuncttrue
\mciteSetBstMidEndSepPunct{\mcitedefaultmidpunct}
{\mcitedefaultendpunct}{\mcitedefaultseppunct}\relax
\EndOfBibitem
\bibitem[Karplus and Petsko(1990)Karplus, and Petsko]{Karplus_1990}
Karplus,~M.; Petsko,~G.~A. \emph{Nature} \textbf{1990}, \emph{347},
  631--639\relax
\mciteBstWouldAddEndPuncttrue
\mciteSetBstMidEndSepPunct{\mcitedefaultmidpunct}
{\mcitedefaultendpunct}{\mcitedefaultseppunct}\relax
\EndOfBibitem
\bibitem[Klepeis \latin{et~al.}(2009)Klepeis, Lindorff-Larsen, Dror, and
  Shaw]{Klepeis_2009}
Klepeis,~J.~L.; Lindorff-Larsen,~K.; Dror,~R.~O.; Shaw,~D.~E. \emph{Current
  Opinion in Structural Biology} \textbf{2009}, \emph{19}, 120--127\relax
\mciteBstWouldAddEndPuncttrue
\mciteSetBstMidEndSepPunct{\mcitedefaultmidpunct}
{\mcitedefaultendpunct}{\mcitedefaultseppunct}\relax
\EndOfBibitem
\bibitem[Tiwary and Walle(2016)Tiwary, and Walle]{tiwary2016review}
Tiwary,~P.; Walle,~A. v.~d. \emph{Multiscale materials modeling for
  nanomechanics} \textbf{2016}, 195--221\relax
\mciteBstWouldAddEndPuncttrue
\mciteSetBstMidEndSepPunct{\mcitedefaultmidpunct}
{\mcitedefaultendpunct}{\mcitedefaultseppunct}\relax
\EndOfBibitem
\bibitem[Henin \latin{et~al.}(2022)Henin, Lelievre, Shirts, Valsson, and
  Delemotte]{He_nin_2022}
Henin,~J.; Lelievre,~T.; Shirts,~M.~R.; Valsson,~O.; Delemotte,~L. \emph{Living
  Journal of Computational Molecular Science} \textbf{2022}, \emph{4}\relax
\mciteBstWouldAddEndPuncttrue
\mciteSetBstMidEndSepPunct{\mcitedefaultmidpunct}
{\mcitedefaultendpunct}{\mcitedefaultseppunct}\relax
\EndOfBibitem
\bibitem[Dickson and Dinner(2010)Dickson, and Dinner]{Dickson_2010}
Dickson,~A.; Dinner,~A.~R. \emph{Annual Review of Physical Chemistry}
  \textbf{2010}, \emph{61}, 441--459\relax
\mciteBstWouldAddEndPuncttrue
\mciteSetBstMidEndSepPunct{\mcitedefaultmidpunct}
{\mcitedefaultendpunct}{\mcitedefaultseppunct}\relax
\EndOfBibitem
\bibitem[Kleiman \latin{et~al.}(2023)Kleiman, Nadeem, and
  Shukla]{kleiman2023adaptive}
Kleiman,~D.~E.; Nadeem,~H.; Shukla,~D. Adaptive Sampling Methods for Molecular
  Dynamics in the Era of Machine Learning. 2023\relax
\mciteBstWouldAddEndPuncttrue
\mciteSetBstMidEndSepPunct{\mcitedefaultmidpunct}
{\mcitedefaultendpunct}{\mcitedefaultseppunct}\relax
\EndOfBibitem
\bibitem[Bonomi \latin{et~al.}(2009)Bonomi, Branduardi, Bussi, Camilloni,
  Provasi, Raiteri, Donadio, Marinelli, Pietrucci, Broglia, and
  Parrinello]{Bonomi2009}
Bonomi,~M.; Branduardi,~D.; Bussi,~G.; Camilloni,~C.; Provasi,~D.; Raiteri,~P.;
  Donadio,~D.; Marinelli,~F.; Pietrucci,~F.; Broglia,~R.~A.; Parrinello,~M.
  \emph{Computer Physics Communications} \textbf{2009}, \emph{180},
  1961--1972\relax
\mciteBstWouldAddEndPuncttrue
\mciteSetBstMidEndSepPunct{\mcitedefaultmidpunct}
{\mcitedefaultendpunct}{\mcitedefaultseppunct}\relax
\EndOfBibitem
\bibitem[Bussi and Laio(2020)Bussi, and Laio]{bussi2020using}
Bussi,~G.; Laio,~A. \emph{Nature Reviews Physics} \textbf{2020}, \emph{2},
  200--212\relax
\mciteBstWouldAddEndPuncttrue
\mciteSetBstMidEndSepPunct{\mcitedefaultmidpunct}
{\mcitedefaultendpunct}{\mcitedefaultseppunct}\relax
\EndOfBibitem
\bibitem[Tiwary and Parrinello(2015)Tiwary, and Parrinello]{Tiwary2015}
Tiwary,~P.; Parrinello,~M. \emph{The Journal of Physical Chemistry B}
  \textbf{2015}, \emph{119}, 736--742\relax
\mciteBstWouldAddEndPuncttrue
\mciteSetBstMidEndSepPunct{\mcitedefaultmidpunct}
{\mcitedefaultendpunct}{\mcitedefaultseppunct}\relax
\EndOfBibitem
\bibitem[Roux(2021)]{stringRoux2021}
Roux,~B. \emph{The Journal of Physical Chemistry A} \textbf{2021}, \emph{125},
  7558--7571\relax
\mciteBstWouldAddEndPuncttrue
\mciteSetBstMidEndSepPunct{\mcitedefaultmidpunct}
{\mcitedefaultendpunct}{\mcitedefaultseppunct}\relax
\EndOfBibitem
\bibitem[Dickson and Brooks(2014)Dickson, and Brooks]{Dickson2014}
Dickson,~A.; Brooks,~C.~L. \emph{The Journal of Physical Chemistry B}
  \textbf{2014}, \emph{118}, 3532--3542\relax
\mciteBstWouldAddEndPuncttrue
\mciteSetBstMidEndSepPunct{\mcitedefaultmidpunct}
{\mcitedefaultendpunct}{\mcitedefaultseppunct}\relax
\EndOfBibitem
\bibitem[Vani \latin{et~al.}(2022)Vani, Weare, and Dinner]{Vani2022}
Vani,~B.~P.; Weare,~J.; Dinner,~A.~R. \emph{The Journal of Chemical Physics}
  \textbf{2022}, \emph{157}\relax
\mciteBstWouldAddEndPuncttrue
\mciteSetBstMidEndSepPunct{\mcitedefaultmidpunct}
{\mcitedefaultendpunct}{\mcitedefaultseppunct}\relax
\EndOfBibitem
\bibitem[Piana and Laio(2008)Piana, and Laio]{PhysRevLett.101.208101}
Piana,~S.; Laio,~A. \emph{Phys. Rev. Lett.} \textbf{2008}, \emph{101},
  208101\relax
\mciteBstWouldAddEndPuncttrue
\mciteSetBstMidEndSepPunct{\mcitedefaultmidpunct}
{\mcitedefaultendpunct}{\mcitedefaultseppunct}\relax
\EndOfBibitem
\bibitem[Ferguson \latin{et~al.}(2011)Ferguson, Panagiotopoulos, Kevrekidis,
  and Debenedetti]{Manifold_debenedetti}
Ferguson,~A.~L.; Panagiotopoulos,~A.~Z.; Kevrekidis,~I.~G.; Debenedetti,~P.~G.
  \emph{Chemical Physics Letters} \textbf{2011}, \emph{509}, 1--11\relax
\mciteBstWouldAddEndPuncttrue
\mciteSetBstMidEndSepPunct{\mcitedefaultmidpunct}
{\mcitedefaultendpunct}{\mcitedefaultseppunct}\relax
\EndOfBibitem
\bibitem[Zwanzig(1961)]{Zwanzig1961}
Zwanzig,~R. \emph{Physical Review} \textbf{1961}, \emph{124}, 983--992\relax
\mciteBstWouldAddEndPuncttrue
\mciteSetBstMidEndSepPunct{\mcitedefaultmidpunct}
{\mcitedefaultendpunct}{\mcitedefaultseppunct}\relax
\EndOfBibitem
\bibitem[Kawasaki(1973)]{Kawasaki1973}
Kawasaki,~K. \emph{Journal of Physics A: Mathematical, Nuclear and General}
  \textbf{1973}, \emph{6}, 1289--1295\relax
\mciteBstWouldAddEndPuncttrue
\mciteSetBstMidEndSepPunct{\mcitedefaultmidpunct}
{\mcitedefaultendpunct}{\mcitedefaultseppunct}\relax
\EndOfBibitem
\bibitem[Antoszewski \latin{et~al.}(2020)Antoszewski, Feng, Vani, Thiede, Hong,
  Weare, Tokmakoff, and Dinner]{Antoszewski2020}
Antoszewski,~A.; Feng,~C.-J.; Vani,~B.~P.; Thiede,~E.~H.; Hong,~L.; Weare,~J.;
  Tokmakoff,~A.; Dinner,~A.~R. \emph{The Journal of Physical Chemistry B}
  \textbf{2020}, \emph{124}, 5571--5587\relax
\mciteBstWouldAddEndPuncttrue
\mciteSetBstMidEndSepPunct{\mcitedefaultmidpunct}
{\mcitedefaultendpunct}{\mcitedefaultseppunct}\relax
\EndOfBibitem
\bibitem[Hong \latin{et~al.}(2018)Hong, Vani, Thiede, Rust, and
  Dinner]{Hong2018}
Hong,~L.; Vani,~B.~P.; Thiede,~E.~H.; Rust,~M.~J.; Dinner,~A.~R.
  \emph{Proceedings of the National Academy of Sciences} \textbf{2018},
  \emph{115}\relax
\mciteBstWouldAddEndPuncttrue
\mciteSetBstMidEndSepPunct{\mcitedefaultmidpunct}
{\mcitedefaultendpunct}{\mcitedefaultseppunct}\relax
\EndOfBibitem
\bibitem[Casasnovas \latin{et~al.}(2017)Casasnovas, Limongelli, Tiwary,
  Carloni, and Parrinello]{Casasnovas2017}
Casasnovas,~R.; Limongelli,~V.; Tiwary,~P.; Carloni,~P.; Parrinello,~M.
  \emph{Journal of the American Chemical Society} \textbf{2017}, \emph{139},
  4780--4788\relax
\mciteBstWouldAddEndPuncttrue
\mciteSetBstMidEndSepPunct{\mcitedefaultmidpunct}
{\mcitedefaultendpunct}{\mcitedefaultseppunct}\relax
\EndOfBibitem
\bibitem[Clark \latin{et~al.}(2016)Clark, Tiwary, Borrelli, Feng, Miller, Abel,
  Friesner, and Berne]{Clark2016}
Clark,~A.~J.; Tiwary,~P.; Borrelli,~K.; Feng,~S.; Miller,~E.~B.; Abel,~R.;
  Friesner,~R.~A.; Berne,~B.~J. \emph{Journal of Chemical Theory and
  Computation} \textbf{2016}, \emph{12}, 2990--2998\relax
\mciteBstWouldAddEndPuncttrue
\mciteSetBstMidEndSepPunct{\mcitedefaultmidpunct}
{\mcitedefaultendpunct}{\mcitedefaultseppunct}\relax
\EndOfBibitem
\bibitem[Mehdi \latin{et~al.}(2023)Mehdi, Smith, Herron, Zou, and
  Tiwary]{mehdi2023enhanced}
Mehdi,~S.; Smith,~Z.; Herron,~L.; Zou,~Z.; Tiwary,~P. Enhanced Sampling with
  Machine Learning: A Review. 2023\relax
\mciteBstWouldAddEndPuncttrue
\mciteSetBstMidEndSepPunct{\mcitedefaultmidpunct}
{\mcitedefaultendpunct}{\mcitedefaultseppunct}\relax
\EndOfBibitem
\bibitem[No{\'{e}} \latin{et~al.}(2020)No{\'{e}}, Tkatchenko, M\"{u}ller, and
  Clementi]{No2020}
No{\'{e}},~F.; Tkatchenko,~A.; M\"{u}ller,~K.-R.; Clementi,~C. \emph{Annual
  Review of Physical Chemistry} \textbf{2020}, \emph{71}, 361--390\relax
\mciteBstWouldAddEndPuncttrue
\mciteSetBstMidEndSepPunct{\mcitedefaultmidpunct}
{\mcitedefaultendpunct}{\mcitedefaultseppunct}\relax
\EndOfBibitem
\bibitem[Rydzewski \latin{et~al.}(2023)Rydzewski, Chen, and
  Valsson]{Rydzewski2023}
Rydzewski,~J.; Chen,~M.; Valsson,~O. \emph{Machine Learning: Science and
  Technology} \textbf{2023}, \relax
\mciteBstWouldAddEndPunctfalse
\mciteSetBstMidEndSepPunct{\mcitedefaultmidpunct}
{}{\mcitedefaultseppunct}\relax
\EndOfBibitem
\bibitem[Meng \latin{et~al.}(2015)Meng, lin Lin, and Roux]{Roux2015}
Meng,~Y.; lin Lin,~Y.; Roux,~B. \emph{The Journal of Physical Chemistry B}
  \textbf{2015}, \emph{119}, 1443--1456\relax
\mciteBstWouldAddEndPuncttrue
\mciteSetBstMidEndSepPunct{\mcitedefaultmidpunct}
{\mcitedefaultendpunct}{\mcitedefaultseppunct}\relax
\EndOfBibitem
\bibitem[Jiang \latin{et~al.}(2022)Jiang, Liu, Liu, Chen, Zheng, and
  Duan]{Jiang2022}
Jiang,~T.; Liu,~Z.; Liu,~W.; Chen,~J.; Zheng,~Z.; Duan,~M. \emph{Journal of
  Chemical Information and Modeling} \textbf{2022}, \emph{62}, 3651--3663\relax
\mciteBstWouldAddEndPuncttrue
\mciteSetBstMidEndSepPunct{\mcitedefaultmidpunct}
{\mcitedefaultendpunct}{\mcitedefaultseppunct}\relax
\EndOfBibitem
\bibitem[Manning \latin{et~al.}(2002)Manning, Whyte, Martinez, Hunter, and
  Sudarsanam]{Manning2002}
Manning,~G.; Whyte,~D.~B.; Martinez,~R.; Hunter,~T.; Sudarsanam,~S.
  \emph{Science} \textbf{2002}, \emph{298}, 1912--1934\relax
\mciteBstWouldAddEndPuncttrue
\mciteSetBstMidEndSepPunct{\mcitedefaultmidpunct}
{\mcitedefaultendpunct}{\mcitedefaultseppunct}\relax
\EndOfBibitem
\bibitem[Bhullar \latin{et~al.}(2018)Bhullar, Lagar{\'{o}}n, McGowan, Parmar,
  Jha, Hubbard, and Rupasinghe]{Bhullar2018}
Bhullar,~K.~S.; Lagar{\'{o}}n,~N.~O.; McGowan,~E.~M.; Parmar,~I.; Jha,~A.;
  Hubbard,~B.~P.; Rupasinghe,~H. P.~V. \emph{Molecular Cancer} \textbf{2018},
  \emph{17}\relax
\mciteBstWouldAddEndPuncttrue
\mciteSetBstMidEndSepPunct{\mcitedefaultmidpunct}
{\mcitedefaultendpunct}{\mcitedefaultseppunct}\relax
\EndOfBibitem
\bibitem[Cohen \latin{et~al.}(2021)Cohen, Cross, and J\"{a}nne]{Kin_DD_21}
Cohen,~P.; Cross,~D.; J\"{a}nne,~P.~A. \emph{Nature Reviews Drug Discovery}
  \textbf{2021}, \emph{20}, 551--569\relax
\mciteBstWouldAddEndPuncttrue
\mciteSetBstMidEndSepPunct{\mcitedefaultmidpunct}
{\mcitedefaultendpunct}{\mcitedefaultseppunct}\relax
\EndOfBibitem
\bibitem[ZHANG and LIU(2002)ZHANG, and LIU]{Kin_pathway}
ZHANG,~W.; LIU,~H.~T. \emph{Cell Research} \textbf{2002}, \emph{12},
  9--18\relax
\mciteBstWouldAddEndPuncttrue
\mciteSetBstMidEndSepPunct{\mcitedefaultmidpunct}
{\mcitedefaultendpunct}{\mcitedefaultseppunct}\relax
\EndOfBibitem
\bibitem[Gadiya and Chakraborty(2018)Gadiya, and Chakraborty]{DDR1_signal}
Gadiya,~M.; Chakraborty,~G. \emph{Cell Adhesion \& Migration} \textbf{2018},
  1--9\relax
\mciteBstWouldAddEndPuncttrue
\mciteSetBstMidEndSepPunct{\mcitedefaultmidpunct}
{\mcitedefaultendpunct}{\mcitedefaultseppunct}\relax
\EndOfBibitem
\bibitem[Hanson \latin{et~al.}(2019)Hanson, Georghiou, Thakur, Miller, Rest,
  Chodera, and Seeliger]{Hanson2019}
Hanson,~S.~M.; Georghiou,~G.; Thakur,~M.~K.; Miller,~W.~T.; Rest,~J.~S.;
  Chodera,~J.~D.; Seeliger,~M.~A. \emph{Cell Chemical Biology} \textbf{2019},
  \emph{26}, 390--399.e5\relax
\mciteBstWouldAddEndPuncttrue
\mciteSetBstMidEndSepPunct{\mcitedefaultmidpunct}
{\mcitedefaultendpunct}{\mcitedefaultseppunct}\relax
\EndOfBibitem
\bibitem[Wang and Tiwary(2021)Wang, and Tiwary]{Dedi}
Wang,~D.; Tiwary,~P. \emph{The Journal of Chemical Physics} \textbf{2021},
  \emph{154}, 134111\relax
\mciteBstWouldAddEndPuncttrue
\mciteSetBstMidEndSepPunct{\mcitedefaultmidpunct}
{\mcitedefaultendpunct}{\mcitedefaultseppunct}\relax
\EndOfBibitem
\bibitem[Roney and Ovchinnikov(2022)Roney, and Ovchinnikov]{Roney2022}
Roney,~J.~P.; Ovchinnikov,~S. \emph{Physical Review Letters} \textbf{2022},
  \emph{129}\relax
\mciteBstWouldAddEndPuncttrue
\mciteSetBstMidEndSepPunct{\mcitedefaultmidpunct}
{\mcitedefaultendpunct}{\mcitedefaultseppunct}\relax
\EndOfBibitem
\bibitem[Dama \latin{et~al.}(2014)Dama, Parrinello, and Voth]{Dama2014}
Dama,~J.~F.; Parrinello,~M.; Voth,~G.~A. \emph{Physical Review Letters}
  \textbf{2014}, \emph{112}\relax
\mciteBstWouldAddEndPuncttrue
\mciteSetBstMidEndSepPunct{\mcitedefaultmidpunct}
{\mcitedefaultendpunct}{\mcitedefaultseppunct}\relax
\EndOfBibitem
\bibitem[Ribeiro and Tiwary(2018)Ribeiro, and Tiwary]{LamimRibeiro2018}
Ribeiro,~J. M.~L.; Tiwary,~P. \emph{Journal of Chemical Theory and Computation}
  \textbf{2018}, \emph{15}, 708--719\relax
\mciteBstWouldAddEndPuncttrue
\mciteSetBstMidEndSepPunct{\mcitedefaultmidpunct}
{\mcitedefaultendpunct}{\mcitedefaultseppunct}\relax
\EndOfBibitem
\bibitem[Mehdi \latin{et~al.}(2022)Mehdi, Wang, Pant, and Tiwary]{Mehdi2022AiB}
Mehdi,~S.; Wang,~D.; Pant,~S.; Tiwary,~P. \emph{Journal of Chemical Theory and
  Computation} \textbf{2022}, \emph{18}, 3231--3238\relax
\mciteBstWouldAddEndPuncttrue
\mciteSetBstMidEndSepPunct{\mcitedefaultmidpunct}
{\mcitedefaultendpunct}{\mcitedefaultseppunct}\relax
\EndOfBibitem
\bibitem[Beyerle \latin{et~al.}(2022)Beyerle, Mehdi, and
  Tiwary]{Beyerle_ent_ener}
Beyerle,~E.~R.; Mehdi,~S.; Tiwary,~P. \emph{The Journal of Physical Chemistry
  B} \textbf{2022}, \emph{126}, 3950--3960\relax
\mciteBstWouldAddEndPuncttrue
\mciteSetBstMidEndSepPunct{\mcitedefaultmidpunct}
{\mcitedefaultendpunct}{\mcitedefaultseppunct}\relax
\EndOfBibitem
\bibitem[Zou \latin{et~al.}(2023)Zou, Beyerle, Tsai, and Tiwary]{Zou2023}
Zou,~Z.; Beyerle,~E.~R.; Tsai,~S.-T.; Tiwary,~P. \emph{Proceedings of the
  National Academy of Sciences} \textbf{2023}, \emph{120}\relax
\mciteBstWouldAddEndPuncttrue
\mciteSetBstMidEndSepPunct{\mcitedefaultmidpunct}
{\mcitedefaultendpunct}{\mcitedefaultseppunct}\relax
\EndOfBibitem
\bibitem[Duan \latin{et~al.}(2003)Duan, Wu, Chowdhury, Lee, Xiong, Zhang, Yang,
  Cieplak, Luo, Lee, \latin{et~al.} others]{amberff03}
others,, \latin{et~al.}  \emph{Journal of computational chemistry}
  \textbf{2003}, \emph{24}, 1999--2012\relax
\mciteBstWouldAddEndPuncttrue
\mciteSetBstMidEndSepPunct{\mcitedefaultmidpunct}
{\mcitedefaultendpunct}{\mcitedefaultseppunct}\relax
\EndOfBibitem
\bibitem[Leimkuhler and Matthews(2013)Leimkuhler, and Matthews]{Leimkuhler2013}
Leimkuhler,~B.; Matthews,~C. \emph{The Journal of Chemical Physics}
  \textbf{2013}, \emph{138}, 174102\relax
\mciteBstWouldAddEndPuncttrue
\mciteSetBstMidEndSepPunct{\mcitedefaultmidpunct}
{\mcitedefaultendpunct}{\mcitedefaultseppunct}\relax
\EndOfBibitem
\bibitem[Eastman \latin{et~al.}(2017)Eastman, Swails, Chodera, McGibbon, Zhao,
  Beauchamp, Wang, Simmonett, Harrigan, Stern, Wiewiora, Brooks, and
  Pande]{Openmm_1}
Eastman,~P.; Swails,~J.; Chodera,~J.~D.; McGibbon,~R.~T.; Zhao,~Y.;
  Beauchamp,~K.~A.; Wang,~L.-P.; Simmonett,~A.~C.; Harrigan,~M.~P.;
  Stern,~C.~D.; Wiewiora,~R.~P.; Brooks,~B.~R.; Pande,~V.~S. \emph{{PLOS}
  Computational Biology} \textbf{2017}, \emph{13}, e1005659\relax
\mciteBstWouldAddEndPuncttrue
\mciteSetBstMidEndSepPunct{\mcitedefaultmidpunct}
{\mcitedefaultendpunct}{\mcitedefaultseppunct}\relax
\EndOfBibitem
\bibitem[Hess \latin{et~al.}(1997)Hess, Bekker, Berendsen, and
  Fraaije]{hess1997lincs}
Hess,~B.; Bekker,~H.; Berendsen,~H.~J.; Fraaije,~J.~G. \emph{Journal of
  computational chemistry} \textbf{1997}, \emph{18}, 1463--1472\relax
\mciteBstWouldAddEndPuncttrue
\mciteSetBstMidEndSepPunct{\mcitedefaultmidpunct}
{\mcitedefaultendpunct}{\mcitedefaultseppunct}\relax
\EndOfBibitem
\bibitem[Darden \latin{et~al.}(1993)Darden, York, and Pedersen]{PME}
Darden,~T.; York,~D.; Pedersen,~L. \emph{The Journal of Chemical Physics}
  \textbf{1993}, \emph{98}, 10089--10092\relax
\mciteBstWouldAddEndPuncttrue
\mciteSetBstMidEndSepPunct{\mcitedefaultmidpunct}
{\mcitedefaultendpunct}{\mcitedefaultseppunct}\relax
\EndOfBibitem
\bibitem[Tiwary \latin{et~al.}(2017)Tiwary, Mondal, and Berne]{Tiwary2017}
Tiwary,~P.; Mondal,~J.; Berne,~B.~J. \emph{Science Advances} \textbf{2017},
  \emph{3}\relax
\mciteBstWouldAddEndPuncttrue
\mciteSetBstMidEndSepPunct{\mcitedefaultmidpunct}
{\mcitedefaultendpunct}{\mcitedefaultseppunct}\relax
\EndOfBibitem
\bibitem[Shekhar \latin{et~al.}(2022)Shekhar, Smith, Seeliger, and
  Tiwary]{Shekhar2022}
Shekhar,~M.; Smith,~Z.; Seeliger,~M.~A.; Tiwary,~P. \emph{Angewandte Chemie
  International Edition} \textbf{2022}, \emph{61}\relax
\mciteBstWouldAddEndPuncttrue
\mciteSetBstMidEndSepPunct{\mcitedefaultmidpunct}
{\mcitedefaultendpunct}{\mcitedefaultseppunct}\relax
\EndOfBibitem
\bibitem[Modi and Dunbrack(2021)Modi, and Dunbrack]{Modi2021}
Modi,~V.; Dunbrack,~R.~L. \emph{Nucleic Acids Research} \textbf{2021},
  \emph{50}, D654--D664\relax
\mciteBstWouldAddEndPuncttrue
\mciteSetBstMidEndSepPunct{\mcitedefaultmidpunct}
{\mcitedefaultendpunct}{\mcitedefaultseppunct}\relax
\EndOfBibitem
\bibitem[Modi and Dunbrack(2019)Modi, and Dunbrack]{Modi2019}
Modi,~V.; Dunbrack,~R.~L. \emph{Proceedings of the National Academy of
  Sciences} \textbf{2019}, \emph{116}, 6818--6827\relax
\mciteBstWouldAddEndPuncttrue
\mciteSetBstMidEndSepPunct{\mcitedefaultmidpunct}
{\mcitedefaultendpunct}{\mcitedefaultseppunct}\relax
\EndOfBibitem
\bibitem[Henzler-Wildman and Kern(2007)Henzler-Wildman, and Kern]{Dyn_Pers}
Henzler-Wildman,~K.; Kern,~D. \emph{Nature} \textbf{2007}, \emph{450},
  964--972\relax
\mciteBstWouldAddEndPuncttrue
\mciteSetBstMidEndSepPunct{\mcitedefaultmidpunct}
{\mcitedefaultendpunct}{\mcitedefaultseppunct}\relax
\EndOfBibitem
\bibitem[Narayan \latin{et~al.}(2020)Narayan, Fathizadeh, Templeton, He,
  Arasteh, Elber, Buchete, and Levy]{Narayan2020}
Narayan,~B.; Fathizadeh,~A.; Templeton,~C.; He,~P.; Arasteh,~S.; Elber,~R.;
  Buchete,~N.-V.; Levy,~R.~M. \emph{Biochimica et Biophysica Acta ({BBA}) -
  General Subjects} \textbf{2020}, \emph{1864}, 129508\relax
\mciteBstWouldAddEndPuncttrue
\mciteSetBstMidEndSepPunct{\mcitedefaultmidpunct}
{\mcitedefaultendpunct}{\mcitedefaultseppunct}\relax
\EndOfBibitem
\bibitem[Saladino and Gervasio(2012)Saladino, and Gervasio]{Saladino2012}
Saladino,~G.; Gervasio,~F.~L. \emph{Current Topics in Medicinal Chemistry}
  \textbf{2012}, \emph{12}, 1889--1895\relax
\mciteBstWouldAddEndPuncttrue
\mciteSetBstMidEndSepPunct{\mcitedefaultmidpunct}
{\mcitedefaultendpunct}{\mcitedefaultseppunct}\relax
\EndOfBibitem
\bibitem[Meng \latin{et~al.}(2018)Meng, Gao, Clawson, Atwell, Russell, Vieth,
  and Roux]{Meng2018}
Meng,~Y.; Gao,~C.; Clawson,~D.~K.; Atwell,~S.; Russell,~M.; Vieth,~M.; Roux,~B.
  \emph{Journal of Chemical Theory and Computation} \textbf{2018}, \emph{14},
  2721--2732\relax
\mciteBstWouldAddEndPuncttrue
\mciteSetBstMidEndSepPunct{\mcitedefaultmidpunct}
{\mcitedefaultendpunct}{\mcitedefaultseppunct}\relax
\EndOfBibitem
\bibitem[Smith \latin{et~al.}(2023)Smith, Strobel, Vani, and Tiwary]{GrASP}
Smith,~Z.; Strobel,~M.; Vani,~B.~P.; Tiwary,~P. \emph{bioRxiv} \textbf{2023},
  \relax
\mciteBstWouldAddEndPunctfalse
\mciteSetBstMidEndSepPunct{\mcitedefaultmidpunct}
{}{\mcitedefaultseppunct}\relax
\EndOfBibitem
\end{mcitethebibliography}
\end{document}